\documentclass{article}

\usepackage{arxiv}
\usepackage{graphicx}
\usepackage[utf8]{inputenc} % allow utf-8 input
\usepackage[T1]{fontenc}    % use 8-bit T1 fonts
\usepackage{hyperref}       % hyperlinks
\usepackage{url}            % simple URL typesetting
\usepackage{booktabs}       % professional-quality tables
\usepackage{amsfonts}       % blackboard math symbols
\usepackage{nicefrac}       % compact symbols for 1/2, etc.
\usepackage{microtype}      % microtypography
\usepackage{lipsum}

\title{Multidimensional Information Assisted Deep Learning Realizing Flexible Recognition of Vortex Beam Modes}

\author{
 Jiale Zhao\\
  Department of Physics\\
  Harbin Institute of Technology\\
  Harbin 150001

   \And
 Zijing Zhang \\
  Department of Physics\\
  Harbin Institute of Technology\\
  Harbin 150001 \\
  \texttt{zhangzijing@hit.edu.cn} \\
  \And
 Yiming Li\\
 College of Information and Electrical Engineering \\
 China Agricultural University\\
 Beijing 100083,\\

  \AND
   Longzhu Cen \\
   Department of Physics\\
   Harbin Institute of Technology\\
   Harbin 150001 \\
}

\begin{document}
\begin{abstract}
\maketitle
Due to countless orthogonal eigenstates, light beam with orbital angular momentum(OAM) has a large potential information capacity. Recently, deep learning has been extensively applied in recognition of OAM mode. However, previous deep learning methods are limited by sign of topological charge (TC) and distance between laser and receiver. In order to further exploit the huge potential of unlimited state space, we proposed a multidimensional information assisted deep learning flexible recognition (MIADLFR) method to make use of both intensity and angular spectrum information to achieve recognition of OAM modes unlimited by sign of TC and distance. With multidimensional feature fusion convolutional neural network (MFFCNN) designed in this paper, we can raise the accuracy of recognition of long distance and strong atmospheric turbulence transmission from 80.1\% to 97.9\%. Also, our method is much less computational expensive, which makes our method more practical.

\end{abstract}


\section{Introduction}
\hspace{0.5cm}Since Allen et al \cite{allen_orbital_1992} recognized that vortex beam with phase structure $exp(il\phi)$  carries OAM $l\hbar $ per photon, vortex beam has been extensively investigated in optical manipulation \cite{paterson_controlled_2001}, imaging \cite{jack_holographic_2009}, optical communication \cite{paterson_atmospheric_2005}. And because of the fact that   can take any integer value, vortex beam has great potential in optical communication \cite{wang_terabit_2012}.

\hspace{0.5cm}In the past 20 years, there are plenty of progress in techniques for sorting OAM modes. Holograms can be used to transform spiral phase structure, thus it can be a mode specific detector \cite{mair_entanglement_2001}. However, this kind of measurement requires a number holograms, which makes it not practical for detecting large number of OAM modes. More efficient sorting can be done with Mach-Zehdner interferometer and a Dove prism in each arm \cite{leach_interferometric_2004}. Theoretically, the efficiency can reach 100\%, but sorting $N$ modes requires $N-1$ cascaded interferometers. Berkhout et al \cite{berkhout_efficient_2010} demonstrated a very successful method for measuring the orbital angular momentum states of light based on log-polar transformation. With this method we can get angular spectrum of vortex beam using two static optical elements. Recently there are also a number of researches improving its resolution \cite{mirhosseini_efficient_2013} \cite{wan_compact_2017} \cite{zheng_improve_2019} \cite{wen_compact_2020} \cite{lightman_miniature_2017}. In fact, gradually-changing-period gratings\cite{dai_measuring_2015} and annular gratings\cite{zheng_measuring_2017} etc. can also sort LG beams, but log-polar transformation seems to be the most intuitive method, as a result of which, we will use log-polar transformation method to extract angular spectrum information in this letter. These methods are useful when light beam is ideal. For vortex beam through atmospheric turbulence, deep learning is usually used to recognize OAM modes. However, there are also some drawbacks in previous deep learning methods for recognizing OAM modes.

\hspace{0.5cm}In recent years, deep learning has been widely applied in computer vision. There are several researches about utilizing deep learning to recognize OAM modes \cite{liu_superhigh-resolution_2019} \cite{wang_efficient_2019}\cite{doster_machine_2017}\cite{liu_deep_2019}\cite{li_joint_2018}\cite{zhai_turbulence_2020}\cite{li_atmospheric_2020}. Zhanwei Liu et al first realized superhigh-resolution recognition of OAM modes with the help of deep learning \cite{liu_superhigh-resolution_2019}. Junmin Liu proposed a deep learning based atmospheric turbulence compensation method \cite{liu_deep_2019}. However, previous deep learning methods seems to have few drawbacks that limit their application.

\hspace{0.5cm}Convolutional neural network (CNN) extracts features from intensity profiles alone in methods proposed before. Nonetheless, as Laguerre-Gaussian (LG) light propagates, the radius of the beam increases, however, the radius of the beam is a quite important feature of LG beam for CNN. Thus, when the training set and testing set of CNN contains LG light propagate different distances, the accuracy of CNN prediction would decrease compared to the same distance case as we will show in the following. Because LG light with the same absolute value of TC however with opposite sign share quite similar intensity profile which is all the information sent into CNN, there is no deep learning method realizing sorting LG light with positive and negative TC efficiently. These drawbacks result from the fact that direct intensity detection cannot provide phase information which gives light beams OAM. However, classically, it is the special phase profile that gives light beam OAM. On the other hand, Angular spectrum information is able to extract part of phase information enough to reveal OAM of light beam directly.

\hspace{0.5cm}In this letter, we proposed and demonstrated a MIADLFR method to remove constrains of preivious deep learning methods mentioned above. MIADLFR method explores both intensity information and spectrum information at the same time with the help of multidimensional feature fusion convolutional neural network (MFFCNN) proposed in this letter. Multidimensional information used together can achieve things impossible for using only intensity information. With multidimensional information, recognition of OAM modes unlimited by distance or sign of TCs can be realized.  What’s more, MIADLFR can also increase the accuracy of recognition of OAM modes, reduce the size of training set and parameters required significantly as we will show in discussion part. At first MFFCNN extract features from intensity information and spectrum information. Then fully connected layers process these two dimensions features and gives prediction. 

%%%%%%%%%%%%%%%%%%%%%%%%%%%%%%%%%%%%%%%%%%%%%%%%%%%%%%%%%%%%%%%%%%%%%%%%%%%%%%%%%%%%%%%%%%%%%%%%5
\section{Theory}
\begin{figure}[ht!]
\centering\includegraphics[width=13cm]{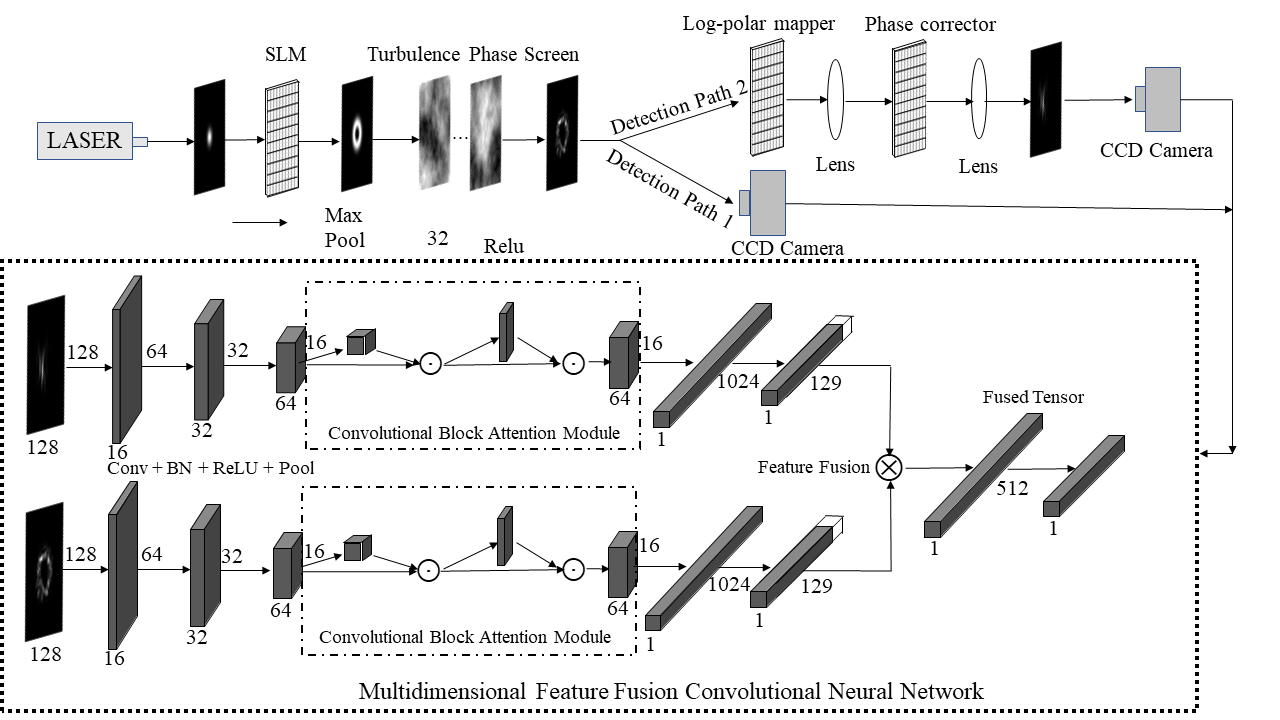}
\caption{The system used for MIADLFR method. SLM is spatial light modulator used to transform gaussian beam to LG beam. Atmospheric turbulence screen is for simulating atmospheric turbulence. Log-polar mapper, phase corrector and lens are used to get angular spectrum information. Parameters of MFFCNN is shown in the schematic diagram.}
\end{figure}

\hspace{0.5cm}The system used in this letter can be seen in Fig.1. Firstly, with the help of spatial light modulator (SLM), Gaussian beam generated by laser can be transformed into LG light. Then LG light gets through atmospheric turbulence simulated by atmospheric turbulence screens. There are two paths for detection in system we used. Through the first path, we get the intensity distribution of LG light after atmospheric turbulence. Through the second path, we get the spectrum information with the help of Log-polar transformation method\cite{berkhout_efficient_2010}. In the end, angular spectrum information and intensity information are sent into MFFCNN together.

\hspace{0.5cm}The complex field becomes LG light $U(z,x,y){{|}_{z=0}}$ after SLM, which carries angular momentum. Then it gets through atmospheric turbulence.

\hspace{0.5cm}For mathematical simplification, treating turbulence as a finite number of discrete atmospheric turbulence screens is a common technique to generate large training sets\cite{liu_deep_2019}. A number of numerical models have been proposed to simulate atmospheric turbulence. In this letter we used model developed by Hill \cite{hill_models_1978} and defined by Andrews \cite{andrews_laser_2005}. The modified von Karman refractive index power spectrum density can be written as:${{\phi }^{2}}(\kappa )=0.49r_{0}^{-5/3}exp(-{{\kappa }^{2}}/\kappa _{m}^{2})/{{({{\kappa }^{2}}+\kappa _{0}^{2})}^{11/6}}$.${{r}_{0}}$ is the effective coherence diameter given by ${{r}_{0}}={{(0.423{{k}^{2}}C_{n}^{2}\Delta z)}^{-3/5}}$. $C_{n}^{2}$ is the atmospheric refractive index structure constant, representing the turbulence intensity. Phase change of atmospheric turbulence screens can be described by $\theta (x,y)=FFT(M\phi (\kappa ))$. $M$ is an $N\times N$ dimensional complex random number array with a mean of 0 and a variance of 1. We might as well set $H=exp(ik\Delta z)\cdot exp(-i(\kappa _{x}^{2}+\kappa _{y}^{2})\Delta z/(2k))$, which represents Fresnel propagation. Then the beam after going through one atmospheric turbulence screen can be described by:
\begin{equation}
U(z,x,y){{|}_{z={{z}_{0}}+\Delta z}}=FF{{T}^{-1}}[H\cdot FFT[exp(i\theta (x,y))\times U(z,x,y){{|}_{z={{z}_{0}}}}]]=\hat{T}U(z,x,y){{|}_{z={{z}_{0}}}}
\end{equation}

\hspace{0.5cm}Thus, intensity profile after $n$ atmospheric turbulence screens can be represented by:
\begin{equation}
{{I}_{1}}(x,y)=|{{\hat{T}}^{n}}U(z,x,y){{|}_{z=0}}{{|}^{2}}
\end{equation}

\hspace{0.5cm}As a result, the intensity distribution detected in the first path is ${{I}_{1}}(x,y)$.

\hspace{0.5cm}As shown in Fig.1. after atmospheric turbulence, one single optical element can be used to achieve the coordinate transformation $(x,y)\mapsto (u,v)$. Here, $v=a\arctan (y/x)$ and $u=-a\ln (\sqrt{{{x}^{2}}+{{y}^{2}}}/b)$. The phase factor of the optical element is given by ${{\phi }_{1}}=2\pi a/(\lambda f)[y\arctan (y/x)-x\ln (\sqrt{{{x}^{2}}+{{y}^{2}}}/b)+x]$. A phase corrector is also needed to eliminate the phase distortion. The phase corrector is given by ${{\phi }_{2}}=-2\pi ab/(\lambda f)exp(-u/a)cos(v/a)$. Detailed explanation can be seen in ref [8]. For simplicity, the light beam after log-polar transformation is represented as:
\begin{equation}
{{U}_{out}}=\hat{L}{{U}_{in}}    
\end{equation}
\hspace{0.5cm}Then, angular spectrum profile sent into CNN can be represented by:
\begin{equation}
{{I}_{2}}(x,y)=|\hat{L}{{\hat{T}}^{n}}U(z,x,y){{|}_{z=0}}{{|}^{2}}    
\end{equation}

\hspace{0.5cm}As a result, the angular spectrum information detected in the second path is ${{I}_{2}}(x,y)$.
As shown in Fig.1. , after log-polar transformation, intensity image and angular spectrum image are sent into two CNN respectively.

\hspace{0.5cm}Deep learning has merged as an important class of artificial intelligence. Recently, deep learning has shown its great power in image classification \cite{he_deep_2016} and CNN is an important tool in deep learning \cite{krizhevsky_imagenet_2017}. CNN is made up with convolutional layers and fully connected layers. Convolutional layers can extract features which are inherently invariant to spatial transformations in images and fully connected layers are able to process information extracted by upstream convolutional layers nonlinearly. After fully connected layers and appropriate activations, we finally get predicted labels.

\hspace{0.5cm}The structure of MFFCNN is shown in Fig.1. What distinguishes MFFCNN from previous CNNs is that the input of MFFCNN contains both intensity and angular spectrum information simultaneously. First, MFFCNN extract features from intensity images and angular spectrum images. In order to ensure the effectiveness of the features, we introduce convolutional block attention module (CBAM)\cite{ferrari_cbam_2018} after convolutional layers, which refines features along channel and spatial axes by inferring their attention maps. The obtained tensors $\mathbf{z}_{1}$, $\mathbf{z}_{2}$ are considered to exist in different feature spaces and represent individual informative meanings. Next, we use a fusion layer \cite{zadeh_tensor_2017} to disentangle dimension specific and cross dimension dynamics by modeling each of them explicitly. The layer is defined as a differentiable outer product between $\mathbf{z}_{1}$ and $\mathbf{z}_{2}$: 
\begin{equation}
\mathbf{z}=\left[\begin{array}{c}
\mathbf{z}_{1} \\
1
\end{array}\right] \otimes\left[\begin{array}{c}
\mathbf{z}_{2} \\
1
\end{array}\right]
\end{equation}
\hspace{0.5cm}Here, $\mathbf{z}$ is the fused tensor; $\otimes$ indicates the outer product between tensors. The extra constant dimension with value 1 generates the dimension specific dynamics and thus $\mathbf{z}$ can be viewed as a 2D square of all possible combinations of two tensor spaces. 

\hspace{0.5cm}Finally, the stacked fully connected layers process the fused tensor $\mathbf{z}$ and give final classification results.

\hspace{0.5cm}Therefore, MFFCNN is used to find a map $f$which can serve as a discriminative boundary among different TCs:
\begin{equation}
\hat{l}=\arg \max f({{I}_{1}}(x,y),{{I}_{2}}(x,y);W)    
\end{equation}
\hspace{0.5cm}Here, $\hat{l}$ is the TC predicted by MFFCNN; $W$ represents trainable parameters in MFFCNN; $f$ is the map to find.

\hspace{0.5cm}In order to find such a map, we need to define loss function to evaluate the difference from prediction and actual TCs. Then through gradient descent we can minimize loss function to make MFFCNN more reliable through mini-batch. The loss function we used can be given by:
\begin{equation}
L=\sum\limits_{i=1}^{M}{\sum\limits_{c=0}^{N-1}{-}}y_{c}^{(i)}\log p_{c}^{(i)}    
\end{equation}

\hspace{0.5cm}Here $M$ is the size of testing set; $N$ is the total number of TC in training set; $y_{c}^{(i)}$ is a binary indicator which takes value 1 if and only if the actual TC of the $ith$ sample of testing set is $c$. $p_{c}^{(i)}$ is the probability of the TC of the $i th$ sample to be $c$ predicted by MFFCNN.

%%%%%%%%%%%%%%%%%%%%%%%%%%%%%%%%%%%%%%%%%%%%%%%%%%%%%%%%%%%%%%%%%%%%%%%%%%%%%%%%%%%%%%%%%%%%%%%%5
\section{Results}
\hspace{0.5cm}The wave length $\lambda $ used in this letter is 532nm. Beam waist ${{w}_{0}}$ of LG light is 0.03m. Size of atmospheric turbulence phase screen is $600\times 600$. Before sending images into MFFCNN images are resized to $128\times 128$. In training sets, there are 600 images for each class of OAM modes. In testing sets, there are 200 images for each class of OAM modes. In log-polar transformation, we used following parameters: $a=0.075/2\pi \mathrm{m}, b=0.075\textrm{m}$. Discussion about number of images for each mode in training set is made below.
\subsection{Recognition of OAM modes for arbitrary distance }
\begin{figure}[h!]
\centering\includegraphics[width=7cm]{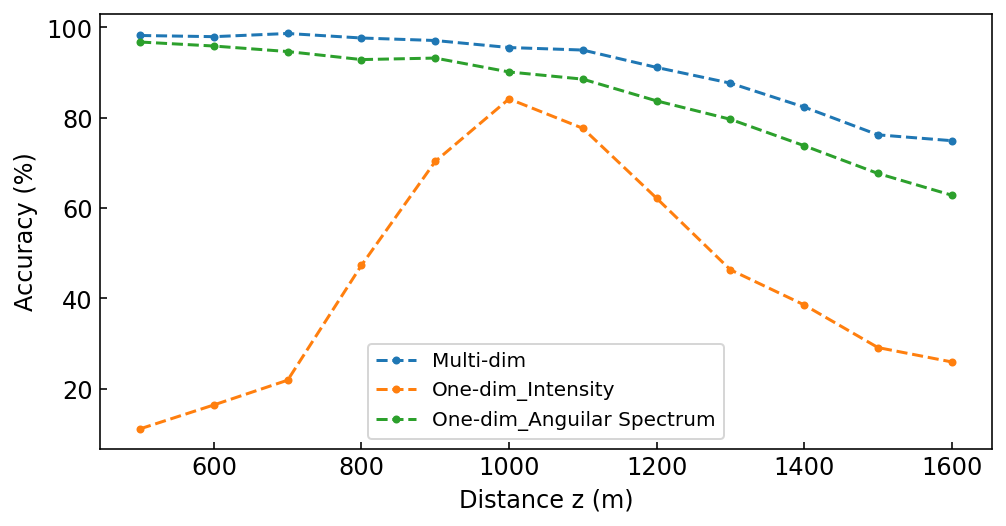}
\caption{Accuracy for of multidimensional and one dimensional recognition of OAM modes propagating various distances through atmospheric turbulence of atmospheric refractive index structure constant $C_{n}^{2}=1\times {{10}^{-14}}$  with TC ranging from $l=1$ to $l=9$.Distance for training set is 1000m.}
\end{figure}
\begin{figure}[ht!]
\centering\includegraphics[width=13cm]{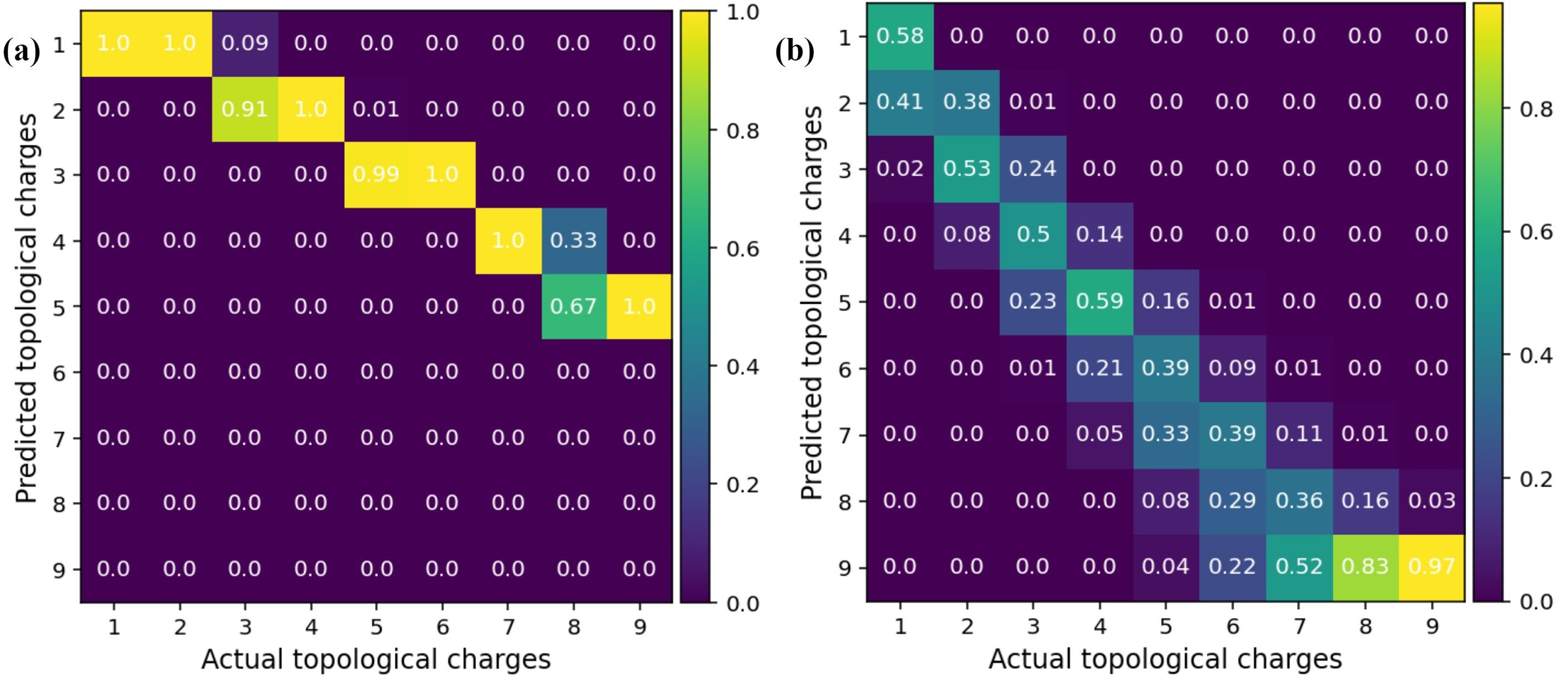}
\caption{(a) Intensity one dimensional recognition accuracy when distance for testing set is 200m. Distance for training set is 1000m. (b) Intensity one-dim recognition accuracy when distance for testing set is 1600m.Distance for training set is 1000m.}
\end{figure}
\hspace{0.5cm}MIADLFR method can be used to recognize OAM modes even when LG light in training set propagate different distance from those in testing set. In order to show that unlike previous deep learning methods(Intensity one-dimensional recognition), recognition achieved by MDIADLFR is robust to change of distance, we first use training set generated in a fixed distance(1000m in our example in Fig.2) just like previous methods and then change the distance for testing sets. As shown in Fig.2., accuracy for traditional methods only reaches peak when distance for training sets and testing set are the same. The accuracy for one dimensional intensity recognition will drop quickly if distance for testing set is different from training set. We believe that this is because as LG light propagate the radius of light beam gets bigger and the radius of light beam is an important feature CNN extracted for mode recognition, as a result of which previous researches are not suitable for OAM modes recognition for distances different from training set. Heat map in Fig.2. can verify our theory. As shown in Fig.3 (a), when the distance for training set is larger than it in testing set, TC predicted by one dimensional CNN is smaller than actual TC. On the contrary if distance for training set is smaller than testing set, TC predicted by one dimensional is larger than actual TC, as we can see in Fig.3. However, for multidimensional recognition, the accuracy is above 90\% for distance less than 1200m and dips slowly as LG light propagate. Accuracy for angular spectrum one dimensional recognition is lower than multidimensional overall. This seems to because features extracted from these two dimensions compensate each other as a result of which accuracy for multidimensional recognition is higher than anyone dimension.

\subsection{Recognition of OAM modes unlimited by sign of TC}
\begin{figure}[ht!]
\centering\includegraphics[width=12cm]{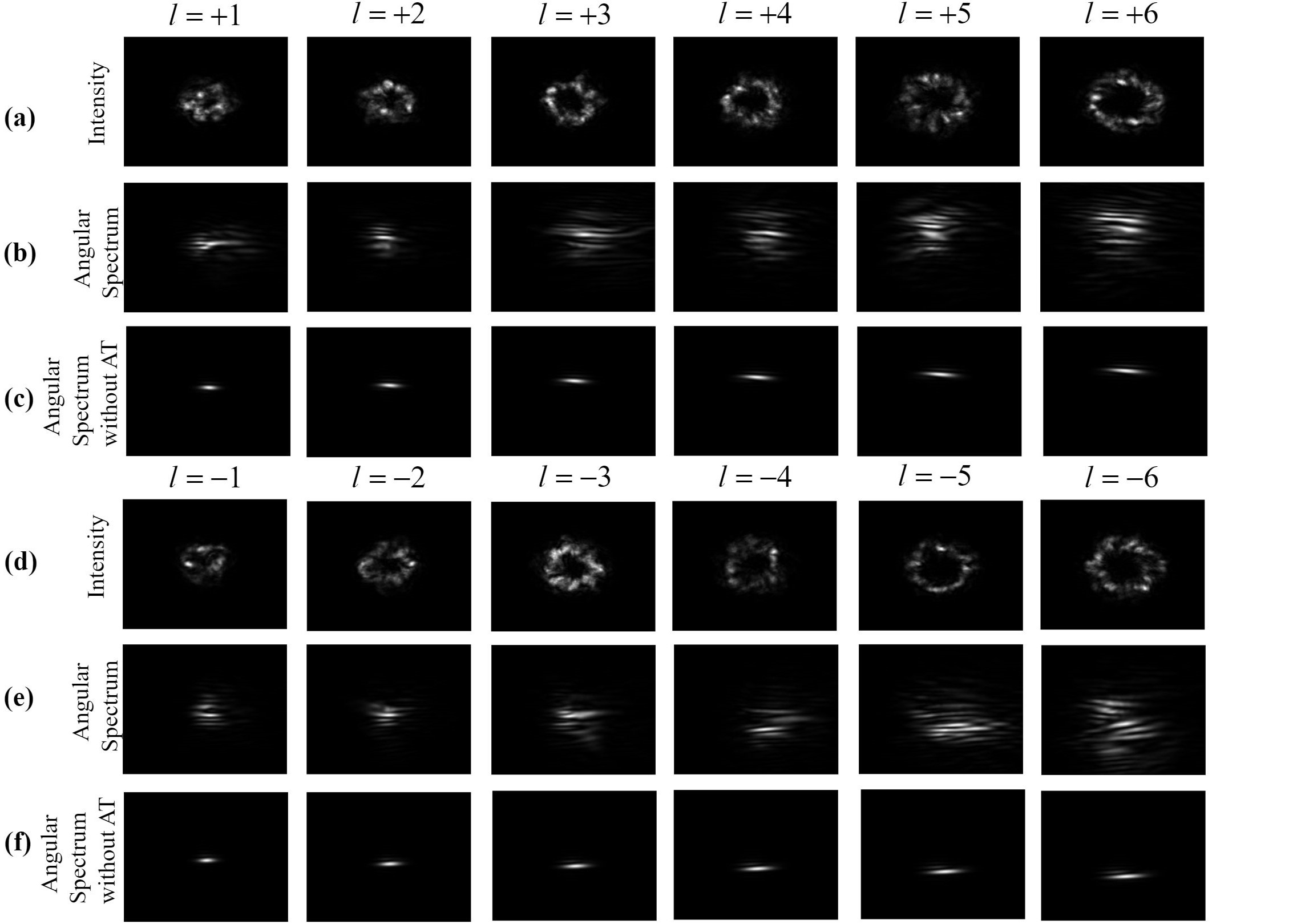}
\caption{(a) (d)Intensity images of LG light propagating 1000m through atmospheric turbulence with $C_{n}^{2}$ value in $1\times {{10}^{-14}}{{m}^{-2/3}}$ carrying various TCs. (b) (e) Angular spectrum images of LG light through atmospheric turbulence with $C_{n}^{2}$ value in $1\times {{10}^{-14}}{{m}^{-2/3}}$ carrying various TCs. (c) (f) Angular spectrum images of LG light without atmospheric turbulence.}
\end{figure}

\begin{figure}[ht!]
\centering\includegraphics[width=12cm]{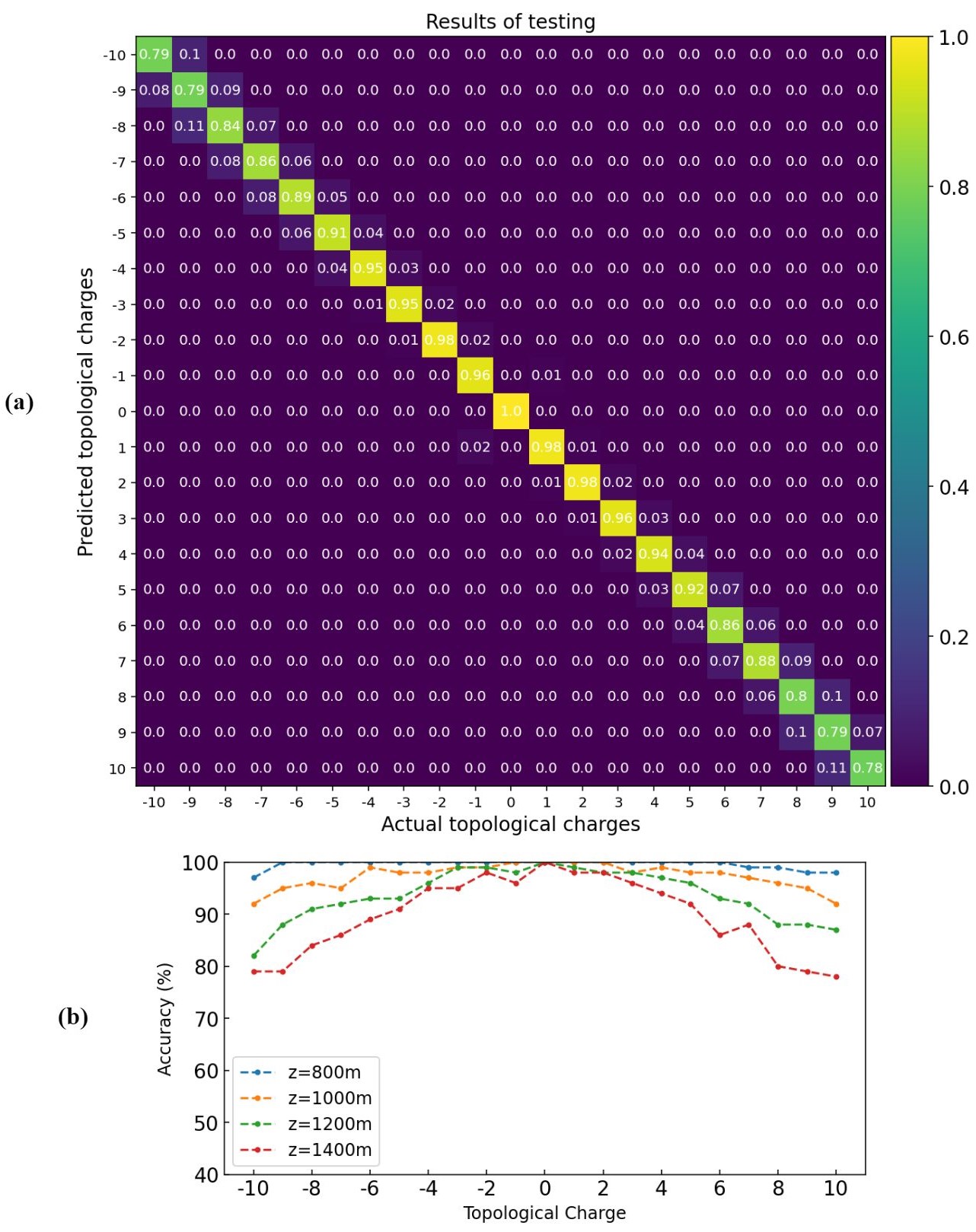}
\caption{(a) Recognition accuracy of OAM modes propagating through atmospheric turbulence with $C_{n}^{2}$ value in$1\times {{10}^{-14}}{{m}^{-2/3}}$ and $l$ ranging from -10 to +10. (b) Accuracy for LG beams with various value of TCs at different distances with $C_{n}^{2}$ equals$1\times {{10}^{-14}}{{m}^{-2/3}}$.}
\end{figure}

\hspace{0.5cm}It is known to all that LG lights with the same absolute value of TC but with opposite sign of TC share the same intensity profile. It can also be seen in Fig.4. (a) (d) that the intensity profiles of OAM modes with the same absolute value of TC but different sign are quite similar. They share the same radius of beam which is a vital feature CNN extracted. Thus, intensity profile is not a quite suitable way to sort OAM modes with positive and negative TCs, as a result of which previous deep learning methods are all limited to positive TC recognition. It can be seen in Fig. 4 (c) (f) that the transverse position in the angular spectrum information detected in path 2 is related to the TC. As shown in Fig. 4 (b) (e), atmospheric turbulence can induce crosstalk between channels especially for adjacent channels. Besides, the crosstalk gets larger as the TC gets larger. 

\hspace{0.5cm}As shown in Fig. 5. (a), MFFCNN is perfectly capable of sorting LG light with positive and negative TC with the same absolute value. As the absolute value of TC grows, the accuracy drops gradually. This is because the crosstalk induced by atmospheric turbulence between channels gets severer as the absolute value of TC grows, as shown in Fig.4 (b) (e), which is consistent with previous researches \cite{anguita_turbulence-induced_2008}.
%%%%%%%%%%%%%%%%%%%%%%%%%%%%%%%%%%%%%%%%%%%%%%%%%%%%%%%%%%%%%%%%%%%%%%%%%%%%%%%%%%%%%%%%%%%%%%%%5
\section{Discussion}

\begin{figure}[ht!]
\centering\includegraphics[width=12cm]{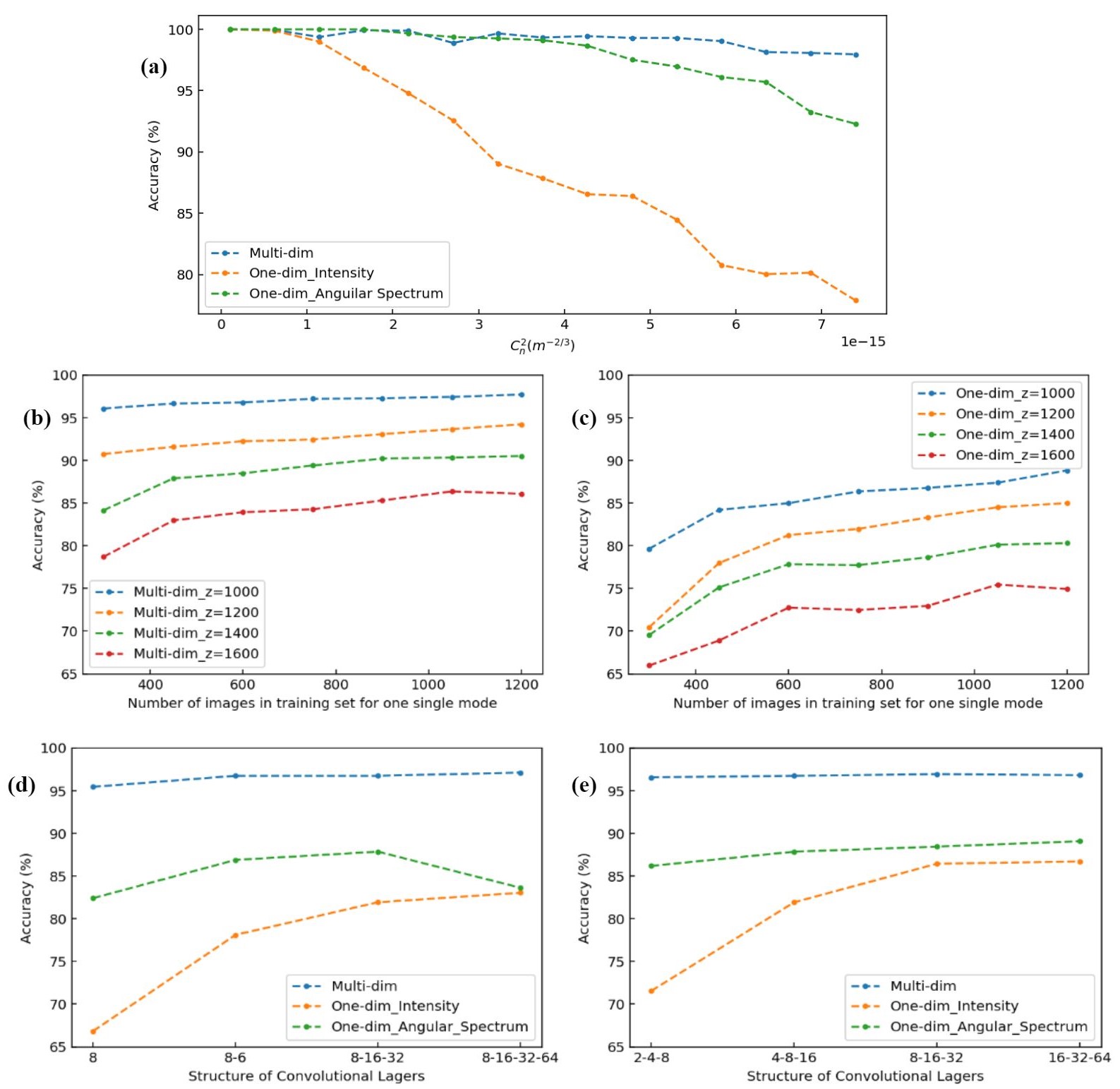}
\caption{ (a) Accuracy of LG light getting through atmospheric turbulence with various $C_{n}^{2}$ when the distance travelled is 1000m. (b) Accuracy of MFFCNN recognition for various number of images in training set for one single mode with $C_{n}^{2}=1\times {{10}^{-14}}$. (c) Accuracy of recognition for various number of images in training set for one single mode with $C_{n}^{2}=1\times {{10}^{-14}}$ using previous methods. (d) Behavior of MFFCNN and one dimensional CNNs with different number of convolutional layers at $z=1000m$ and $C_{n}^{2}=1\times {{10}^{-14}}$. $x$ coordinates represent the structure of convolutional layers. For example, ’16-32’ means that there are two convolutional layers with 16 and 32 channels respectively. (e) Behavior of MFFCNN and one dimensional CNNs with different number of channels at $z=1000m$ and $C_{n}^{2}=1\times {{10}^{-14}}$.}
\end{figure}

\subsection{Accuracy for various strength of atmospheric turbulence}
\hspace{0.5cm}It is shown in Fig.6. (a) that the accuracy of intensity one dimensional recognition falls since the structure constant of refractive index reaches $1\times {{10}^{-15}}$. This is because for intensity one dimensional recognition, the intensity profile especially the radius of light beam changes greatly when the atmospheric turbulence is strong. For angular spectrum one dimensional recognition, when the atmospheric turbulence gets stronger the crosstalk between channels tend to be symmetric, thus accuracy slides much slower than intensity one dimensional recognition. As for multidimensional recognition, MFFCNN extract features from both intensity profile and angular spectrum so that recognition of OAM modes is less affected by atmospheric turbulence and the accuracy is over 97\% when $C_{n}^{2}$ reaches$7\times {{10}^{-15}}$.
\subsection{Size of training set}
\hspace{0.5cm}As is known to all that larger training set usually result in higher accuracy of course below certain upper limit. Generating training set usually take a lot of time and it takes more time to train a model with larger training set. As a result, we need to make a trade-off about the size of training set. In most occasions, usually choose points after which the accuracy rises quite slow as the size of training set grows. We might as well refer the size for each class as converge point (CP) in this letter.
It can be seen in Fig.6. (b) (c) that the longer LG light propagated that is to say the severer LG light is affected, the larger training set is needed. When LG light propagates 1000m and 1200m, the CP point is 300 for MFFCNN. Meanwhile, the CP points are 600 and 750 for $z=1400$ and $z=1600$ respectively. This result is reasonable, because we usually need larger training set to find best map for classification when the input gets more noise. As for previous methods, CP points are 600 for various distances as shown in Fig. 6. (c). In a nut shell, MFFCNN only need a half of size for previous method which can make better use of the potential advantage of large OAM state space.

\subsection{Complexity}
\hspace{0.5cm}From Fig.6.(d) (e), we can see that MFFCNN does not need a complex CNN structure to reach a very high accuracy. In Fig.6.(d), we can also see that MFFCNN can get 95\% accuracy with only one convolutional layer, meanwhile accuracy for one dimensional CNN is only 65\%. As we can see one dimensional CNN needs three layers to reach a point after which more complex CNN structure would not raise the accuracy significantly. We believe this is because angular spectrum dimension information compensates features that need complex CNN to extract from intensity dimension. However, features that cannot be extracted from angular spectrum dimension do not need complex CNN to extract from intensity dimension to extract. As shown in Fig.6.(e), intensity one dimensional recognition needs larger number of channels to reach a stable point, which also means that intensity one dimensional recognition is computationally expensive. What’s more, with the help of intensity dimension information, accuracy for MFFCNN is 10\% higher than angular spectrum one dimensional recognition the whole time. To sum up, MFFCNN is much less computationally demanding.
%%%%%%%%%%%%%%%%%%%%%%%%%%%%%%%%%%%%%%%%%%%%%%%%%%%%%%%%%%%%%%%%%%%%%%%%%%%%%%%%%%%%%%%%%%%%%%%%%
\section{Conclusion}
\hspace{0.5cm}In this letter, we proposed MIADLFR method realize flexible recognition of OAM modes. MIADLFR method can extract features from both intensity and angular spectrum thus it can realize things previous deep learning methods fail to achieve, and raise accuracy of prediction remarkably. Such method can recognize OAM modes propagating different distances and sort OAM modes with positive and negative TCs. When atmospheric turbulence refractive index reaches $7\times {{10}^{-5}}$ accuracies are $80.1\%$, $93.2\%$, $97.9\%$ for intensity one dimensional, angular spectrum, multidimensional recognition respectively. MFFCNN only need 300 images for each mode in training set to reach a relatively stable point for accuracy. Meanwhile, previous methods need at least 600 images. Smaller training set needed would reduce workload for both generating training set and training remarkably, thus make it easier to show the potential advantage of large OAM state space. What’s more, MFFCNN needs much simpler CNN structure than previous methods, which makes it much faster to train and recognize OAM modes and thus more practical.

\hspace{0.5cm}MIADLFR method proposed in this letter can not only be used in OAM modes recognition, it can be applied in other problems especially for those require multiple inputs. For some problems, collection of training set is technically or computationally expensive, in which case MFFCNN can also come to use.

\section{Funding}
This work is supported by National Natural Science Foundation of China (Grant No. 61701139 and No. 62075049).
\section{Disclosures}
The authors declare that they have no known competing financial interests or personal relationships that could have appeared to influence the work reported in the paper entitled “Multidimensional Information Assisted Deep Learning Realizing Flexible Recognition of Vortex Beam Modes”.

\bibliographystyle{unsrt}

\end{document}